%Paper: hep-th/9312112
%From: huebschm@gat.univ-lille1.fr (huebschmann johannes)
%Date: Tue, 14 Dec 93 10:58:38 +0100

%This paper is written in AMSTeX
\documentstyle{amsppt}
\magnification=1200

\hoffset=-0.5pc
\vsize=57.2truepc
\hsize=38truepc
\nologo
\spaceskip=.5em plus.25em minus.20em

 \define\atiboton{1}
 \define\atibottw{2}
 \define\bergever{3}
\define\bottone{4}
\define\botshust{5}
\define\goldmone{6}
\define\singula{7}
\define\singulat{8}
\define\topology{9}
\define\smooth{10}
\define\poisson{11}
\define\locpois{12}
\define\singulth{13}
\define\jeffrone{14}
\define\jeffrtwo{15}
\define\karshone{16}
\define\kirwafou{17}
\define\kirwaboo{18}
\define\maclaboo{19}
\define\narshtwo{20}
\define\narashed{21}
\define\sjamlerm{22}
\define\weinstwe{23}

\noindent{\bf PUB. IRMA, Lille - 1993\newline\noindent
Vol. 33, N$^{\roman o}$ \, X}\newline\noindent
hep-th/9312112
\bigskip

\noindent
\topmatter
\title
Symplectic and Poisson structures
of certain moduli spaces
\endtitle
\author Johannes Huebschmann
\endauthor
\affil
Universit\'e des Sciences et Technologie
de Lille
\\
U. F. R. de Math\'ematiques
\\
F-59 655 VILLENEUVE D'ASCQ Cedex, France
\\
Huebschm\@GAT.UNIV-LILLE1.FR
\endaffil
\abstract{
Let  $\pi$
be the fundamental group
of a closed surface and $G$ a Lie group with a biinvariant metric,
not necessarily positive definite.
It is shown that a certain construction
due to A. Weinstein
relying on
techniques  from equivariant
cohomology
may be refined
so as to yield
(i) a
symplectic structure on
a certain smooth manifold $\Cal M(\Cal P,G)$
containing the space
$\roman{Hom}(\pi,G)$
of homomorphisms
and, furthermore,
(ii)
a hamiltonian  $G$-action
on $\Cal M(\Cal P,G)$
preserving the symplectic structure,
with momentum mapping $\mu \colon \Cal M(\Cal P,G) \to g^*$,
in such a way that the reduced space
equals the space
$\roman{Rep}(\pi,G)$ of representations.
Our approach is somewhat more general in that it
also applies to twisted
moduli spaces;
in particular,
it yields
the {\smc Narasimhan-Seshadri}
moduli spaces of semistable
holomorphic vector
bundles
by {\it symplectic reduction
in finite dimensions\/}.
This implies that,
when the group $G$ is compact,
such a twisted moduli space
inherits a structure of {\it stratified symplectic space\/},
and that
the strata of these twisted moduli spaces have
finite symplectic volume.}
\endabstract
\date{November 30, 1993}
\enddate
\endkeywords
\endsubjclass

\endtopmatter
\document

\beginsection Introduction

Let  $\pi$
be the fundamental group
of a closed surface and $G$ a Lie group with a biinvariant metric,
not necessarily positive definite.
The symplectic structure on
the smooth part of
(the components of)
the moduli space
$\roman{Rep}(\pi,G) = \roman{Hom}(\pi,G)\big / G$
\cite\atibottw,
\cite\goldmone\
has recently been obtained
by A. Weinstein
by entirely finite dimensional
techniques \cite\weinstwe.
Weinstein constructed
a closed equivariant 2-form on
(the smooth part of)
$\roman{Hom}(\pi,G)$
and showed by techniques from equivariant
cohomology \cite\atiboton\
that this 2-form descends to
(the non-singular part of)
$\roman{Rep}(\pi,G)$.
In this paper we refine Weinstein's method by showing
that in fact it yields
(i) a
symplectic structure on
a certain smooth manifold $\Cal M(\Cal P,G)$
containing
$\roman{Hom}(\pi,G)$
(as a deformation retract)
and, furthermore,
(ii)
a hamiltonian  $G$-action
on $\Cal M(\Cal P,G)$
preserving the symplectic structure,
with momentum mapping $\mu \colon \Cal M(\Cal P,G) \to g^*$,
in such a way that the reduced space
equals
$\roman{Rep}(\pi,G)$.
Our approach is somewhat more general in that it
also applies to twisted
moduli spaces, see Section 6 below.
In particular,
it yields
the {\smc Narasimhan-Seshadri}
\cite\narashed\
moduli spaces of semistable
holomorphic vector
bundles
by {\it symplectic reduction, applied to
a smooth finite dimensional symplectic
manifold with a hamiltonian action of
the unitary group\/}, which is {\it finite dimensional\/}.
Our result, apart from being interesting in its own right,
reveals some interesting and attractive geometric properties of these
twisted moduli spaces.
Firstly, it implies that,
when the group $G$ is compact,
with reference to the
decomposition of orbit types,
such a twisted moduli space
inherits a structure of {\it stratified symplectic space\/}
in the sense of \cite\sjamlerm;
in particular, this yields a
description of the behaviour
the symplectic
or more generally Poisson structure
of these moduli spaces
in their singularities (in the appropriate sense);
see our paper \cite\locpois\ where
the Poisson geometry has been worked out
explicitly in some special cases.
Secondly our results implies that
the strata of these twisted moduli spaces have
finite symplectic volume.
See Section 7 below for details.
The finiteness of the symplectic volume
of the
moduli spaces
of semistable
holomorphic vector
bundles has been obtained also by {\smc Narasihman\/}
[unpublished]
by an argument involving the Quillen metric.
Moreover it is likely that
the momentum mapping which we found
can be studied further
by means of the well known convexity
results, cf. \cite\kirwafou\
and the literature there.
We also expect
that {\smc Kirwan's} methods
\cite\kirwaboo\
or suitable extensions thereof
become applicable so that more calculations
in the cohomology of these moduli spaces can be done.
\smallskip
Our approach in fact
applies to arbitrary
groups $\pi$ with a finite presentation and yields
then somewhat more general results.
It may be viewed as a step towards the
\lq\lq grand unified theory\rq\rq\
searched for by A. Weinstein
in that, cf. Lemma 1 below, it explains  the occurence
of the form $\vartheta$ in \cite\weinstwe\
which there was put in \lq\lq by hand\rq\rq.
\smallskip
An extension of our construction yields other cohomology classes
from the characteristic ring, in fact, it also applies
to integral cohomology via the cohomology of the
classifying space of $G$. We shall make this precise
at another occasion.
\smallskip
For the flat case,
the space $\roman{Rep}(\pi,G)$
has been obtained
as the reduced space
of a certain smooth finite dimensional
symplectic manifold
with hamiltonian $G$-action
also by L. Jeffrey \cite\jeffrone.
Her construction involves gauge theory techniques.
After completion of the paper
I learnt that some of the same results
have been obtained by her independently \cite\jeffrtwo.
\smallskip
I am much indebted to J. Stasheff,
A. Weinstein, and Dong Yan
for a number of comments which helped
clarify the exposition, and to
R. Sjamaar for
explaining me the requisite technical details for the proof
in \cite\sjamlerm\
of the finiteness of the symplectic volumes of the strata
of a reduced space under appropriate circumstances; in fact the proof
of (7.2) below crucially relies thereupon.

\beginsection 1. Equivariant Maurer-Cartan calculus

Let $G$ be a
Lie group,
with Lie algebra $g$,
and let
$\cdot$ be an
adjoint action invariant symmetric bilinear form
on $g$, not necessarily definite nor positive.
We use the notation and terminology
in \cite\weinstwe.
\smallskip
For ease of exposition we recall some notions
from equivariant de Rham cohomology \cite\atiboton, \cite\bergever.
Let $M$ be
a $G$-manifold $M$, and
consider
its
bi-differential graded
algebra
$(\Omega_G^{*,*}(M);d, \delta_G)$
of $G$-equivariant forms; here
the Lie algebra $g$ is concentrated in degree two,
and $\Omega_G^{2j,k}(M)$ refers to the
$G$-equivariant
homogeneous degree
$j$
polynomial maps
on $g$ with values in the usual degree $k$ de Rham forms $\Omega^{k}(M)$.
The usual
de Rham differential on $M$ is $G$-invariant and hence induces an operator
\newline\noindent
--- $d\colon \Omega_G^{2j,k}(M) \to \Omega_G^{2j,k+1}(M)$,
for every $k,j \geq 0$;
\newline\noindent
the additional {\it equivariant\/} operator
\newline\noindent
--- $\delta_G\colon \Omega_G^{2j,k}(M) \to \Omega_G^{2j+2,k-1}(M)$,
for $k\geq 1$ and $j \geq 0$,
\newline\noindent
assigns
to a $G$-invariant polynomial map
$\alpha \colon g \to \Omega^k(M)$
the $G$-invariant polynomial map
$\delta_G(\alpha) \colon g \to \Omega^{k-1}(M)$
which is given by
$$
(\delta_G(\alpha))(X) = - i_{X_M}(\alpha(X)),
$$
where $X_M$ denotes the vector field
given by the Lie algebra element
$X$ acting on $M$.
It is clear that $dd = 0$ and
a straightforward contraction and Lie derivative
calculation shows that
$$
d \delta_G + \delta_G d = 0,\quad
\delta_G \delta_G = 0.
\tag 1
$$
Consequently the operator
$$
d + \delta_G
$$
is a differential on the total
object of $\Omega_G^{*,*}(M)$.
It is clear that
$(\Omega_G^{*,*}(\cdot);d, \delta_G)$
is a functor in the appropriate sense.
This is all we need to know about equivariant
cohomology; for our purposes it is merely an excellent tool
to say something concisely which might have been technically more
involved in some other language.
We only mention that
we could as well have
defined
$(\delta_G(\alpha))(X)$
by
$i_{X_M}(\alpha(X))$
but we have chosen the above operator
to arrive at the same formulas as in \cite\weinstwe.
\smallskip
For every $q \geq 1$, the direct product $G^q$ of $q$ factors of $G$
will be considered as a $G$-manifold, with $G$-action by inner automorphisms
on each copy of $G$.
In the usual way these fit together to a simplicial manifold
$NG$
having $G^q$ in degree $q$, and
$NG$
inherits an obvious structure of
simplicial $G$-manifold \cite\bottone, \cite\botshust.
Application of the functor
$(\Omega_G^{*,*}(\cdot);d, \delta_G)$
to $NG$
yields a cosimplicial
bi-differential graded
algebra
$(\Omega_G^{*,*}(NG);d, \delta_G)$ --- we did not indicate
the cosimplicial structure explicitly.
Its realization yields
the {\it equivariant bar de Rham\/}
 tricomplex
$$
\left(\Omega_G^{*,*}(G^*); \delta, d, \delta_G\right).
$$
Here
$\delta$
refers to the (inhomogeneous) bar complex operator.
Recall (p. 116 of \cite\maclaboo) it is given by the formula
$$
\aligned
\delta(f)(a_0,\dots,a_q)
&= (-1)^{q+1}\left[
f(a_1,\dots,a_q) \right.
\\
& \quad +
\sum_{j=1}^{q-1}
(-1)^j
f(a_0,\dots,a_{j-1}a_j,\dots,a_q)
\\
& \left.\quad +
(-1)^q
f(a_0,\dots,a_{q-1})\right].
\endaligned
$$
The corresponding
(inhomogeneous) bar complex operator
$\delta\colon \Omega_G^{*,*}(G^{q}) \to \Omega_G^{*,*}(G^{q+1})$
looks like
$$
\delta f = (-1)^{q+1} \sum_{j=0}^q (-1)^j \delta_j f;
$$
here
$f$ refers to a de Rham form on $G^q$,
$\delta_0 f$ and
$\delta_q f$
are the forms on
$G^{q+1}$
arising from
pulling back $f$ from
$G^{q}$
via the projection mappings from
$G^{q+1}$
to $G^{q}$
forgetting the first and last copy of $G$, respectively,
while for $1 \leq j \leq q-1$,
$\delta_j f$ is
the form on $G^{q+1}$
arising from pulling back $f$ from
$G^{q}$
via the map
which multiplies the $(j-1)$th with the $j$th
variable and fixes the rest.
By construction,
the operator $\delta$
is compatible with the other structure, and hence
the operator
$d_G$,
which on elements of total {\it form degree\/} $p$ is given by
$$
d_G= \delta + (-1)^p (d+ \delta_G),
$$
is a differential on the totalization for
$\left(\Omega_G^{*,*}(G^*); \delta, d, \delta_G\right)$.
The
resulting complex
is
the {\it total  equivariant bar de Rham complex}.
The above sign
$(-1)^{q+1}$ does not occur in
\cite\weinstwe;
according
to the Eilenberg-Koszul convention
it is the formally appropriate sign.
Since the only explicit instance
of the operator $\delta$ in \cite\weinstwe\
refers to a situation where $q=1$, see (2) and (4) below,
our sign convention does not affect the formulas.
\smallskip
For intelligibility we now
recall briefly
the equivariant Maurer-Cartan calculus
from \cite\weinstwe:
Denote by $\omega$ the $g$-valued left-invariant 1-form
on $G$
which maps each
tangent vector to the left invariant vector field having that value.
The corresponding right invariant form will be denoted by $\overline\omega$.
Recall that the {\it triple product\/}
$(x,y,z) \mapsto [x,y]\cdot z$
yields
an alternating trilinear form  on $g$;
for our purposes, the appropriate form is
$$
\tau(x,y,z) = \frac 1{2} [x,y]\cdot z,
\quad
x,y,z \in g;
$$
it is closed for the
usual Cartan-Chevalley-Eilenberg
Lie algebra cohomology operator.
Its left translate is a closed invariant 3-form
$\lambda$ on $G$
which by the way coincides with the right
translate of $\tau$.
Notice
$$
\lambda = \frac 1{12} [\omega,\omega] \cdot\omega.
$$
At the risk of making a mountain of a molehill
we recall that the present conventions
entail that, for arbitrary left invariant vector fields $X$ and $Y$
on $G$, we have
$[\omega,\omega](X,Y) = 2[X,Y]$ whence the sum of the three
requisite shuffles yields
$$
([\omega,\omega] \cdot\omega)(X,Y,Z)
= 6[X,Y]\cdot Z,
$$
for arbitrary left invariant vector fields $X,Y,Z$,
and thence the left translate $\lambda$ of $\tau$
looks like stated.
Next, if $\alpha$ is any differential form on $G$,
we denote by $\alpha_j$ the pullback of $\alpha$ to
$G\times G$
by the projection $p_j$
to the j'th component.
Let
$$
\Omega = \frac 12 \omega_1 \cdot \overline \omega_2.
$$
This is an alternating 2-form on $G \times G$.
Here the convention is that
$$
(\omega_1 \cdot \overline \omega_2)
(U,V) =
\omega_1 (U)\cdot \overline \omega_2(V)
-
\overline \omega_2 (U)\cdot \omega_1(V),
$$
that is,
there is {\it no\/} factor of $\frac 12$,
and
the pairing of arbitrary forms will likewise
involve shuffles instead of all permutations.
This
is the convention coming into play in \cite\weinstwe;
it is forced by
the form of the Maurer-Cartan equations
in \cite\weinstwe\  and, furthermore, by
the identity (4) below, and we keep this convention
to arrive at the same formulas as in \cite\weinstwe.
Finally, consider the element
$\vartheta \in \Omega_G^{2,1}(G)$,
that is, the linear $G$-invariant map
$\vartheta \colon g \to \Omega^{1}(G)$
whose adjoint
$\vartheta^{\flat}$, viewed as a $g^*$-valued 1-form on
$G$, amounts to
$\frac 12(\omega + \overline \omega)$, combined with
the adjoint
$g \to g^*$
of the given 2-form;
thus, when we view $X\in g$ as a constant
$g$-valued $0$-form on $G$,
$$
\vartheta(X) = \frac 12 X \cdot (\omega + \overline \omega).
$$
We note that the forms $\lambda,\Omega, \vartheta$
are exactly half the forms denoted by the same symbols in
\cite\weinstwe.
The reason why we work here with
half the forms in \cite\weinstwe\
will become clear in Sections 4 and 5 below.
In fact, the form $\Omega$ will eventually yield
the symplectic structures on moduli spaces.
\smallskip
With these preparations out of the way, the equivariant
Maurer-Cartan calculus
amounts to the following relations:
$$
\alignat 2
d\Omega &= \delta \lambda
\quad &&\text{\cite \weinstwe\ (3.3)},
\tag2
\\
\delta \Omega &= 0
\quad &&\text{\cite \weinstwe\ (3.4)},
\tag3
\\
\delta_G \Omega &= -\delta \vartheta
\quad &&\text{\cite \weinstwe\ (4.4)},
\tag4
\\
d \lambda &= 0
\quad &&\text{\cite \weinstwe\ (3.1)},
\tag5
\\
\delta_G \lambda &= d\vartheta
\quad &&\text{\cite \weinstwe\ (4.1)},
\tag6
\\
\delta_G \vartheta &= 0
\quad &&\text{\cite \weinstwe\ (4.3)}.
\tag7
\endalignat
$$
These identities say that
(i) the form
$\Omega - \lambda$
is an equivariant closed form
({\it not\/} equivariantly closed)
of (total) degree 4
and that
(ii) the form
$Q_4 = \Omega - \lambda + \vartheta$
is an equivariantly closed element of (total) degree 4 in the total
complex of the equivariant bar de Rham complex
$(\Omega_G^{*,*}(G^*); d,\delta, \delta_G)$,
cf. \cite\weinstwe\ (4.5).
\smallskip
We conclude this Section with another consequence of these
relations which will be crucial in the sequel.
Let
$h \colon \Omega^*(g) \to \Omega^{*-1}(g)$
be an adjoint action invariant  homotopy operator,
so that the operator $dh + hd$ equals the identity
on $\Omega^*(g)$.
For example we could take
the usual homotopy operator
given by
integration
of forms along linear paths, but
this will not be important for us
since we shall only need the mentioned formal
properties of $h$.
Write $\rho = \exp^* (\lambda) \in \Omega^3(g)$
and let $\beta = h (\rho)\in \Omega^2(g)$.
Then it is obvious that
$$
d \beta = \rho
\tag 8
$$
in the {\it de Rham} complex
of $g$.
Moreover,
since $h$
is
equivariant, so is
$\beta$,
that is,
$\beta \in \Omega_G^{0,2}(g)$, with
reference to the adjoint representation.
As usual we write
$\exp \colon g \to G$
for the exponential mapping
and $\exp^*$
for the induced map
from $\Omega_G^{*,*}(G)$ to $\Omega_G^{*,*}(g)$.

\proclaim{Lemma 1}
There is a smooth $G$-equivariant
map
$\psi \colon g \to g^*$, uniquely
determined by the requirement
that $\psi(0) = 0$,
whose adjoint
$\psi^{\sharp}\colon g \to  \Omega^0(g) = C^{\infty}(g)$, that is,
$\psi^{\sharp} \in \Omega_G^{2,0}(g)$,
satisfies
$$
d\psi^{\sharp} =
\exp^* (\vartheta) + \delta_G(\beta).
$$
\endproclaim

To prove this we observe at first that
(1) and (6) imply
$$
d(\exp^* (\vartheta) + \delta_G\beta)
= 0,
$$
that is,
$\exp^* (\vartheta) + \delta_G\beta$
is a cocycle for the de Rham operator $d$.
In fact,
$$
\aligned
d(\exp^* (\vartheta) + \delta_G\beta)
&=\exp^* (d\vartheta) + d\delta_G\beta
\\
&
=\exp^*(d\vartheta) - \delta_G (d \beta)
\\
&=\exp^*(d\vartheta) - \delta_G (\rho)
\\
&=\exp^*(d\vartheta - \delta_G (\lambda)) = 0.
\endaligned
$$
Hence
the smooth $G$-equivariant map $\psi$
whose adjoint
is given by
$$
\psi^{\sharp} = h \circ (\exp^* (\vartheta) + \delta_G(\beta))
\colon g \to C^{\infty}(g)
$$
has the asserted property.
It is manifestly unique up to a constant. \qed
\smallskip

\proclaim{Corollary}
The value of the derivative of $\psi$ at
an arbitrary point of the centre
of $g$ equals
the adjoint
$g \to g^*$
of the given 2-form
on $g$.
Consequently
the restriction of $\psi$ to the centre
equals itself
the adjoint of this 2-form.
\endproclaim

\demo{Proof}
Let $z$ be a point of the centre of $g$.
By Lemma 1, the derivative amounts to
the sum of
$\vartheta^{\flat}_z \colon g \to g^*$,
the adjoint
$\vartheta^{\flat}$
of $\vartheta$,
viewed as a $g^*$-valued 1-form on $G$,
at $\roman{exp} (z) \in G$,
and
$(\delta_G\beta)^{\flat}_z\colon g \to g^*$,
the adjoint
$(\delta_G\beta)^{\flat}$
of
$\delta_G\beta \in \Omega_G^{2,1}(g)$,
viewed as a $g^*$-valued 1-form on $g$,
at $z \in g$.
Since
$\vartheta^{\flat}$, viewed as a $g^*$-valued 1-form on
$G$, amounts to $\frac 12(\omega + \overline \omega)$, combined with
the adjoint
$g \to g^*$
of the given 2-form on $g$,
the derivative
$\vartheta^{\flat}_z$
equals the adjoint
$g \to g^*$ of this 2-form.
To understand the other term,
recall that the vector field $X_G$
for the adjoint representation
amounts to
the sum $X + \overline X$, where
$\overline X$ is the right invariant vector field on $G$
having at the identity of $G$ the same value as $X \in g$.
By construction,
$\delta_G\beta$
is the  equivariant linear
map
$\delta_G\beta \colon g \to \Omega^{1}(g)$
given by
$$
((\delta_G\beta)(X))(Y) = - \beta (X_G,Y),
$$
for $X,Y \in g$, where
$X_G$
and $Y$
refer to the vector fields on $g$
arising from pulling
back via the exponential mapping the vector fields on $G$
denoted by the same symbols;
in particular $Y$ should {\it not\/} be viewed as
the vector field on $g$
arising from the usual identification of
a vector space with its tangent space
at any point. In fact, the latter
amounts to the {\it canonical\/} parallelization on $g$,
viewed as a vector space, whereas the vector field
$Y$ arises from the parallelization
given by {\it left translation\/} on $G$.
Thus all told, we obtain the formula
$$
((\delta_G\beta)(X))(Y) = \beta (\overline X -X,Y)
=\beta (\overline X ,Y)- \beta (X,Y),
$$
for $X,Y \in g$.
However, at a point of the centre, the two vector
fields
$\overline X$
and
$X$
coincide whence
$(\delta_G\beta)^{\flat}_z\colon g \to g^*$ is zero. \qed
\enddemo

\smallskip\noindent{\smc Remark.}
Suppose $h$ is the standard
homotopy operator
given by
integration
of forms along linear paths in $g$.
Then it may be shown that
$\delta_G h+ h\delta_G$
is zero and a calculation
of L. Jeffrey \cite\jeffrtwo\
shows that
the map $\psi$ amounts to the adjoint of
the given 2-form on all of $g$.
In fact, in view of (6),
$$
\align
d(h \circ \roman{exp}^*(\vartheta))
&= (\roman{Id} - hd) \roman{exp}^*(\vartheta)
\\
&= \roman{exp}^*(\vartheta)
- h \roman{exp}^*(\delta_G(\lambda))
\\
&= \roman{exp}^*(\vartheta)
+\delta_G h \roman{exp}^*(\lambda))
\\
&= \roman{exp}^*(\vartheta)
+\delta_G (\beta),
\endalign
$$
whence we may  take
$\psi = h \circ \roman{exp}^*(\vartheta)$.
Given $x$ and $y$ in $g$, we then have
$$
\align
\psi(x) \cdot y
&= \int_0^1(\roman{exp}^*(\vartheta))_{tx}(x) \cdot y
\\
&= \int_0^1\vartheta_{\roman {exp}(tx)}(\roman{exp}_*x)\cdot y
\\
&= \frac 12 \int_0^1(\omega + \overline \omega)_{\roman {exp}(tx)}
(\roman{exp}_*x)\cdot y
\\&= x \cdot y,
\endalign
$$
since
along the
1-parameter subgroup generated by $x$
(i) the vector field $\roman{exp}_*x$
coincides with $x$
and (ii)
$\overline \omega(x)$ equals
$\omega(x) = x$.

%\beginsection 2. Forms on representation spaces

\medskip\noindent{\bf 2. Forms on representation spaces}
\smallskip\noindent
Let $\Pi$ be a group;
we denote
the chain and cochain complexes
of its inhomogeneous {\it reduced\/} normalized bar
resolution
$B\Pi$
\cite\maclaboo\
by $(C_*(\Pi),\partial)$ and
$(C^*(\Pi),\delta)$, respectively.
For
$p \geq 1$, consider the evaluation map
$$
E \colon \Pi^p \times \roman{Hom}(\Pi,G) @>>> G^p
\tag9
$$
and the induced maps
$$
E^*\colon
\Omega_G^{*,*}(G^p) @>>>
C^p(\Pi) \otimes
\Omega_G^{*,*}(\roman{Hom}(\Pi,G)).
\tag10
$$
They assemble to a morphism
$$
\left(\Omega_G^{*,*}(G^*); d,\delta, \delta_G\right)
@>>>
(C^*(\Pi),\delta) \otimes
\left(\Omega_G^{*,*}(\roman{Hom}(\Pi,G)); d,\delta_G\right)
\tag11
$$
of tricomplexes;
in a given tridegree $(i,j,k)$, it looks like
$$
\Omega_G^{i,j}(G^k)
@>>>
C^k(\Pi) \otimes
\Omega_G^{i,j}(\roman{Hom}(\Pi,G)).
$$
Pairing with chains in
$C_*(\Pi)$, we obtain
the chain map
$$
\aligned
\langle\cdot,\cdot\rangle
\colon
(C_*(\Pi),\partial)\otimes (C^*(\Pi),\delta) \otimes
&\left(\Omega_G^{*,*}(\roman{Hom}(\Pi,G)); d,\delta_G\right)
\\
&@>>>
\left(\Omega_G^{*,*}(\roman{Hom}(\Pi,G)); d,\delta_G\right)
\endaligned
\tag12
$$
which produces
equivariant forms
on $\roman{Hom}(\Pi,G)$, suitably interpreted in the singularities
of $\roman{Hom}(\Pi,G)$; actually, the problem of singularities
will  {\it not\/} occur below.
In particular, for a 2-chain $c \in C_2(\Pi)$, let
$$
\omega_c = \langle c, E^*Q_4\rangle.
\tag13
$$
Since (11) is a morphism of tricomplexes, this amounts to
$$
\omega_c =\langle c, E^*\Omega\rangle,
\tag14
$$
and, cf. (1) and (3),
$$
\align
d\omega_c &=
d\langle c, E^*\Omega\rangle
=
\langle c, E^*d\Omega\rangle
=
\langle c, E^*\delta\lambda\rangle
=
\langle \partial c, E^*\lambda
\rangle,
\tag15
\\
\delta_G\omega_c &=
\delta_G\langle c, E^*\Omega\rangle
=
\langle c, E^*\delta_G \Omega\rangle
=
\langle c,- E^*\delta \vartheta \rangle
=
-\langle \partial c, E^*\vartheta\rangle.
\tag16
\endalign
$$
In particular, when $c$ is closed,
$\omega_c$
is closed {\it and\/} equivariantly closed;
moreover,
when $c = \partial b$,
$$
\langle \partial b, E^*\Omega \rangle =
\langle b, E^*\delta \Omega \rangle = 0,
$$
cf. (2), whence
the assignment of
$\omega_c$
to $c$ yields a homomorphism from
$\roman H_2(\Pi)$ to the space of closed and
equivariantly closed equivariant 2-forms
on $\roman{Hom}(\Pi,G)$, cf. \cite\weinstwe\ (5.1).
For
$\kappa \in \roman H_2(\Pi)$, we therefore write
$\omega_\kappa \in \Omega^2(\roman{Hom}(\Pi,G))$
for its image, to indicate that this form depends merely
on the cohomology class.
\smallskip
Let
$$
\Cal P = \langle x_1,\dots,x_n; r_1,\dots,r_m\rangle
$$
be a presentation of a group
$\pi$, write $F$ for the free group on the generators
and $N$ for the normal closure
of the relators, so that $\pi = F/N$.
The
choice of generators identifies
$\roman{Hom}(F,G)$ with $G^n$ and, furthermore, the
space $\roman{Hom}(\pi,G)$ with the pre-image
of
the identity element in $G^m$, for the word map
$$
r =(r_1,\dots,r_m) \colon G^n @>>> G^m
$$
induced by the relators.
Let $O\subseteq g$ be the open $G$-invariant subset of $g$
where the exponential mapping
from $g$ to $G$
is regular;
notice that $O$ contains the centre of $g$.
Write
$\Cal H(F,G)$
for the
space determined by the requirement that
a pull back diagram
$$
\CD
\Cal H(\Cal P,G)
@>r>> O^m
\\
@V{\eta}VV
@VV{\roman{exp}}V
\\
\roman{Hom}(F,G)
@>>r> G^m
\endCD
\tag 17
$$
results, where the induced map
from $\Cal H(\Cal P,G)$ to $O^m$ is still denoted by $r$.
The space
$\Cal H(\Cal P,G)$
is a smooth manifold
and the induced map
$\eta$ from
$\Cal H(\Cal P,G)$
to
$\roman{Hom}(F,G) =G^n$
is a smooth codimension zero immersion whence
$\Cal H(\Cal P,G)$ has the same dimension as $G^n$;
moreover
the above injection
of $\roman{Hom}(\pi,G)$
into
$\roman{Hom}(F,G)$
induces a canonical
injection
of $\roman{Hom}(\pi,G)$
into
$\Cal H(\Cal P,G)$,
and in this way
$\roman{Hom}(\pi,G)$
will be viewed as a subspace of
$\Cal H(\Cal P,G)$;
when we restrict to a suitable $G$-invariant ball in
$O^m$, we obtain even an open
submanifold of $G^n$ but this will not be important for us.
\smallskip
Let $c$ be a 2-chain of $F$ whose image $c_\pi$ in
$C_2(\pi)$ under the canonical map is closed, and write
$\kappa=[c_\pi] \in \roman H_2(\pi)$ for its class.
We can then apply the above construction, to $\Pi = F$
and $\Pi= \pi$.
With
$\Pi = F$
we obtain a 2-form $\omega_c$ on
$\roman{Hom}(F,G)= G^n$
which, by naturality, on $\roman{Hom}(\pi,G)$
restricts  to the closed 2-form
$\omega_{\kappa}$
(in the appropriate sense)
corresponding to
$\kappa \in \roman H_2(\pi)$.
In \cite\weinstwe, A. Weinstein
raised the question whether
$\omega_{\kappa}$
admits a closed extension
to
$\roman{Hom}(F,G)= G^n$.
Here is a possible answer to this question, with $G^n$ replaced
by the space $\Cal H(\Cal P,G)$.

\proclaim{Theorem 1}
For every class $\kappa \in \roman H_2(\pi)$,
there is a 2-chain $c \in C_2(F)$
and an equivariant 2-form $B$ on
$g^m$ determined by
$c$ and $\Cal P$
so that
$\omega_{c,\Cal P} = \eta^*(\omega_c) - r^*B$
is
a closed $G$-equivariant 2-form on
$\Cal H(\Cal P,G)$
extending $\omega_{\kappa}$.
\endproclaim

\noindent{\smc Remark 1.}
The theorem does not assert that the extension is equivariantly closed.
In fact, in general it will {\it not\/} be equivariantly closed.
An equivariantly closed extension will be constructed
in the next Section.

\proclaim{Lemma 2}
For every $\kappa \in \roman H_2(\pi)$,
there is a
2-chain
$c \in C_2(F)$
whose image
$c_\pi \in  C_2(\pi)$
represents
$\kappa \in \roman H_2(\pi)$
and whose boundary
$\partial c$ looks like
$$
\partial c = \sum \nu_j r_j,\quad
\nu_j \in \bold Z.
$$
Moreover,
$\partial c$ then represents $\kappa$,
the second homology group
$\roman H_2(\pi)$ being identified with the kernel of the
induced map
from $N \big / [F,N]$ to
$F \big / [F,F]$
by means of the Schur-Hopf formula.
\endproclaim

\demo{Proof}
Pick a 2-chain
$c \in C_2(F)$
whose image
$c_\pi \in  C_2(\pi)$
represents
$\kappa \in \roman H_2(\pi)$.
The boundary $\partial c$ is a finite linear combination
$$
\partial c = \sum \nu_j w_j,\quad
\nu_j \in \bold Z,\ w_j \in F.
$$
We assert that for a suitable choice,
each $w_j$ lies in $N$. In fact,
$\partial c$ lies in the kernel of the induced map
from $C_1(F)$ to  $C_1(\pi)$.
Since
$C_1(F)$ and  $C_1(\pi)$
are the free abelian groups generated by $F^*= F \setminus 1$ and
$\pi^* = \pi \setminus 1$,
the kernel
of the induced map
from $C_1(F)$ to  $C_1(\pi)$
is generated by
the elements of $N^*$ together with elements of the kind
$[wn] -[w]$, for $w\in F$ and $n \in N$.
However, for $w\in F$ and $n \in N$,
$$
\partial [w|n] = [n] -[wn] + [w],
$$
and $[w|n] \in C_2(F)$ goes to zero in
$C_2(\pi)$. Hence a suitable modification of $c$
if necessary yields a 2-chain $c' \in C_2(F)$
so that
(i) $\partial c'$ is a sum of terms $\mu_k w_k$ where
each $w_k \in N$
and (ii) whose image
$c'_\pi \in  C_2(\pi)$
still represents
$\kappa \in \roman H_2(\pi)$.
Thus we may assume that the boundary of $c$ involves
only elements of $N$.
\smallskip
Next we show that we may
in fact pick $c$ in
the asserted way.
In fact,
each $w \in N$ is a product of conjugates $yr_j y^{-1}$
and $zr^{-1}_k z^{-1},\, y,z \in F$.
However,
$$
\partial [u|v] = [v] - [uv] + [u],\quad
\partial [u|u^{-1}] =    [u^{-1}] + [u],
$$
whence adding suitable terms to $c$ if necessary, we may assume
$\partial c$ is of the kind
$$
\partial c = \sum \nu_j y_jr_jy^{-1}_j,\quad
\nu_j \in \bold Z;
$$
notice adding these terms
does not change the image $c_{\pi} \in C_2(\pi)$.
Finally,
$$
\partial[xv|x^{-1}] = [x^{-1}] - [xvx^{-1}] +[xv],
\quad
\partial[x^{-1}|xv] = [xv] -[v] +[x^{-1}],
$$
whence adding further terms to $c$ if necessary, we
arrive at a 2-chain in $C_2(F)$
of the asserted kind.
The statement involving the Schur-Hopf formula
is then obvious.
\qed
\enddemo

\demo{Proof of Theorem 1}
Let $c$ be a
2-chain
for $F$
whose image
$c_\pi \in  C_2(\pi)$
represents
$\kappa \in \roman H_2(\pi)$
and whose boundary
$\partial c$ looks like
$$
\partial c = \sum \nu_j r_j,\quad
\nu_j \in \bold Z.
$$
By Lemma 2, such a $c$ exists.
For $j = 1,\dots,m$, let
$p_j \colon G^m \to G$
be the projection onto the j'th factor,
let $\lambda_j = p_j^*(\lambda)\in \Omega^3(G^m)$,
and let
${
\Lambda = \sum \nu_j \lambda_j \in\Omega^3(G^m).
}$
This is a closed equivariant 3-form and,
since
$E \colon F \times \roman{Hom}(F,G) @>>> G$
maps $(w,\phi)$ to the result of application of the
word map for $w$ to $(\phi(x_1),\dots,\phi(x_n))$,
$$
\langle \partial c, E^*\lambda\rangle = r^*\Lambda
$$
whence
$$
d\omega_c = r^*\Lambda \in \Omega^3(G^n).
\tag 18
$$
For $j = 1,\dots,m$, let
$\beta_j = p_j^*(\beta)\in \Omega^2(g^m)$,
and let
${
B =  \sum \nu_j \beta_j \in\Omega^2(g^m)
}$
so that
$$
dB =  \sum \nu_j d\beta_j
= \sum \nu_j \roman{exp}^*(\lambda_j)
=\roman{exp}^*(\Lambda)\in\Omega^3(g^m).
$$
Then on
$\Cal H(\Cal P,G)$ we have the identity
${
d\eta^*(\omega_c) = r^*dB,
}$
that is,
$$
d(\eta^*(\omega_c) - r^*B)= 0.
$$
The form
$\eta^*(\omega_c) - r^*B $ is the closed equivariant extension
of $\omega_\kappa$
we are
looking for. \qed
\enddemo

\smallskip\noindent
{\smc Remark 2.}
The 2-form $\omega_c$ may be closed on a space larger than
$\roman{Hom}(\pi,G)$.
In fact,
the 3-form $\lambda$ is identically zero on
the centre $Z$ of $G$.
Hence,
in view of the identity (18),
the naturality of the construction implies that,
the space $G^n$ being identified with $\roman{Hom}(F,G)$
by means of the choice of generators,
the
2-form $\omega_c$
is closed
on the subspace
of $\roman{Hom}(F,G)$
consisting of
homomorphisms $\phi$ from $F$ to $G$ having the property that
$\phi(r_j)$ lies in $Z$, for $1 \leq j \leq m$.

\beginsection 3. The equivariantly closed extension

Let  $\kappa \in \roman H_2(\pi)$,
and let $c \in C_2(F)$
be a 2-chain and
$B$
the corresponding equivariant 2-form on
$g^m$ so that
$\omega_{c,\Cal P} = \eta^*(\omega_c) - r^*B$
is an equivariant closed 2-form
$\Cal H(\Cal P,G)$
extending $\omega_\kappa$, as in Theorem 1.

\proclaim{Theorem 2}
There is a smooth equivariant map
$
\mu \colon
\Cal H(\Cal P,G)
@>>> g^*
$
whose adjoint
$\mu^\sharp \colon g \to C^{\infty}(\Cal H(\Cal P,G))$
satisfies
the identity
$$
\delta_G (\omega_{c,\Cal P})
= d \mu^{\sharp}
$$
on $\Cal H(\Cal P,G)$.
Consequently
$\omega_{c,\Cal P}
-\mu^\sharp$
is an equivariantly closed
form
in
\linebreak
$(\Omega_G^{*,*}(\Cal H(\Cal P,G));d,\delta_G)$
of total degree 2 extending $\omega_\kappa$.
\endproclaim

Thus, cf. \cite\atiboton\ and
what is said in Section 5 below,
$\mu$ is a momentum mapping for
the 2-form $\omega_{c,\Cal P}$ on
$\Cal H(\Cal P,G)$, with reference to the
obvious $G$-action, except
that
$\omega_{c,\Cal P}$
is not necessarily non-degenerate.
\smallskip

\demo{Proof}
For $j = 1,\dots,m$, let
$\vartheta_j = p_j^*(\vartheta)\in \Omega_G^{2,1}(G^m)$, and let
$
\Theta = \sum \nu_j \vartheta_j  \in \Omega_G^{2,1}(G^m)
$
so that
$$
\langle \partial c, E^*\vartheta\rangle = r^*\Theta \in \Omega_G^{2,1}(G^n).
$$
In view of (16),
$$
\align
\delta_G (\omega_{c,\Cal P})
&=
\delta_G (\eta^*(\omega_c) - r^*B)
\\
&=
\eta^*(\delta_G \omega_c) - r^*\delta_G B
\\
&=
-\eta^*(\langle \partial c, E^*\vartheta\rangle)
 - r^*\delta_G B
\\
&= -r^*(\exp^*(\Theta) +\delta_G B)
\in \Omega_G^{2,1}(\Cal H(\Cal P,G)).
\endalign
$$
For $j = 1,\dots,m$, write
$\psi_j \colon g^m \to g^*$
for the composite of the projection onto the j'th factor
with $\psi$, so that its adjoint
$\psi_j^{\sharp}\colon g \to  \Omega^0(g^m) = C^{\infty}(g^m)$, that is,
$\psi_j^{\sharp} \in \Omega_G^{2,0}(g^m)$,
satisfies
$$
d\psi_j^{\sharp} =
(\exp^*(\vartheta_j) +\delta_G \beta_j).
$$
Let
$$
\Psi = \sum \nu_j \psi_j \colon g^m \to g^*;
$$
this is a smooth map.
By Lemma 1, its adjoint
$\Psi^{\sharp} \in \Omega_G^{2,0}(g^m)$
satisfies
$$
d\Psi^{\sharp} =
\sum \nu_j (\exp^*(\vartheta_j) +\delta_G \beta_j)
=\exp^* (\Theta) + \delta_G B \in \Omega_G^{2,1}(g^m).
$$
Hence on
$\Cal H(\Cal P,G)$ we have the identity
$$
\delta_G (\omega_{c,\Cal P})
=-r^*(d\Psi^{\sharp}) = -d (r^*\Psi^{\sharp})
$$
which, with
$
\mu = -\Psi \circ r \colon
\Cal H(\Cal P,G)
@>>> g^*,
$
looks like
$$
\delta_G (\omega_{c,\Cal P})
= d \mu^\sharp
$$
as asserted. \qed
\enddemo

\beginsection 4. Examination of the equivariantly closed 2-form

Let $\phi \in \roman{Hom}(F,G) = G^n$
and suppose that each $\phi(r_j)$ lies in the centre of $G$.
Then the
composite of $\phi$ with the adjoint representation of $G$
induces a
structure of a $\pi$-module
on $g$. We write
$g_{\phi}$ for
$g$, with
the resulting $\pi$-module structure.
By means of the {\it left} Fox calculus, the presentation
$\Cal P$
determines a chain complex
$$
\bold C(\Cal P, g_{\phi})\colon
\roman C^0(\Cal P, g_{\phi})
@>{\delta_{\phi}^0}>>
\roman C^1(\Cal P, g_{\phi})
@>{\delta_{\phi}^1}>>
\roman C^2(\Cal P, g_{\phi})
\tag4.1
$$
computing the group cohomology $\roman H^*(\pi,g_{\phi})$
in degrees 0 and 1; we only mention that
there are canonical isomorphisms
$$
\roman C^0(\Cal P, g_{\phi}) \cong g,
\quad
\roman C^1(\Cal P, g_{\phi}) \cong g^n,
\quad
\roman C^2(\Cal P, g_{\phi}) \cong g^m .
$$
See e.~g. our paper
\cite\smooth\ where
the details are given for the {\it right} Fox calculus.
To explain the geometric significance
of this chain complex,
denote by $\alpha_\phi$
the smooth map
from $G$  to $\roman{Hom}(F,G)$
which assigns $x \phi x^{-1}$ to $x \in G$,
and write
$R_\phi\colon g^n \to  \roman T_{\phi} \roman{Hom}(F,G)$
and
$R_{r\phi}\colon g^m \to  \roman T_{r\phi}G^m$
for the corresponding operations of right translation.
The tangent maps
$\roman T_e\alpha_{\phi}$
and $\roman T_{\phi}r$
make commutative the diagram
$$
\CD
\roman T_eG
@>\roman T_e\alpha_{\phi}>>
\roman T_{\phi} \roman{Hom}(F,G)
@>{\roman T_{\phi} r}>>
\roman T_{r (\phi)}G^m
\\
@A{\roman{Id}}AA
@A{\roman R_{\phi}}AA
@A{\roman R_{r(\phi)}}AA
\\
g
@>>{\delta^0_{\phi}}>
g^n
@>>{\delta^1_{\phi}}>
g^m.
\endCD
\tag4.2
$$
We work here with right translation
since this yields formulas consistent with
existing literature on representation spaces,
cf. e.~g. \cite\narshtwo.
In our paper
\cite\smooth\
we worked with left translation since there the relationship
with principal bundles is made explicit, and the usual convention
in the literature is to have the structure
group  of a principal bundle act on the right.
\smallskip
The
2-form $\cdot$ on $g$
and the homology class $\kappa$
determine the alternating
2-form
$$
\omega_{\kappa,\cdot,\phi}
\colon
\roman H^1(\pi,g_{\phi})
\otimes
\roman H^1(\pi,g_{\phi})
@>{\cup}>>
\roman H^2(\pi,\bold R)
@>{\cap \kappa}>> \bold R
\tag4.3
$$
on
$\roman H^1(\pi,g_{\phi})$.
In view of the commutativity of (4.2),
right translation
identifies the
kernel of
the derivative
$\roman T_{\phi} r$
with the kernel
of the coboundary operator
$\delta_{\phi}^1$
from
$\roman C^1(\Cal P, g_{\phi})$
to
$\roman C^2(\Cal P, g_{\phi})$,
that is, with the
vector space
$\roman Z^1(\pi,g_{\phi})$
of $g_{\phi}$-valued 1-cocycles of $\pi$;
this space does {\it not\/}
depend on a specific presentation $\Cal P$, whence the notation.
We note that $\roman C^1(\Cal P, g_{\phi}) =
Z^1(F,g_{\phi})$,
the space of $g_{\phi}$-valued 1-cocycles for $F$.
Our present goal is to prove the following.

\proclaim{Theorem 4.4}
Right translation identifies the
restriction of the
2-form $\omega_{c}$
to the kernel of
the derivative
$\roman T_{\phi} r$
with
the 2-form
on $\roman Z^1(\pi,g_{\phi})$
obtained as the
composite of
$\omega_{\kappa,\cdot,\phi}$
with the projection
from
$\roman Z^1(\pi,g_{\phi})$
to
$\roman H^1(\pi,g_{\phi})$.
\endproclaim

\smallskip
While this is essentially contained in \cite\weinstwe, see also \cite\karshone,
we give a complete proof since our argument will clarify the role of the
factor $\frac 12$ in the definition of the forms $\lambda,\Omega,\vartheta$.
\smallskip
For a group $\Pi$, as usual we
write $\beta(\Pi)$ for the
{\it unreduced\/} normalized inhomogeneous
bar resolution;
the {\it reduced\/} normalized inhomogeneous
bar resolution $B\Pi$ arises from
$\beta(\Pi)$
by dividing out the $\Pi$-action.

\proclaim{Lemma 4.5}
The cup-pairing
$
\cup \colon
Z^1(F,g_{\phi})
\otimes
Z^1(F,g_{\phi})
@>>>
C^2(F,\bold R)
$
with reference to the 2-form $\cdot$\  on $g$
is given by the formula
$$
(u \cup v)[x|y] = u(x) \cdot (\roman{Ad}(\phi(x))v(y)),
\quad
u,v \in Z^1(F,g_{\phi}),\
[x|y] \in B_2(\Pi).
$$
\endproclaim

\demo{Proof}
Let $\Pi$ be an arbitrary discrete group.
Recall from \cite\maclaboo\ (VIII.9 Ex. 1, p. 248)
that
the relevant term
of the {\it Alexander-Whitney} diagonal map
$\Delta \colon \beta(\Pi) \to \beta(\Pi) \otimes \beta(\Pi)$
is given by the formula
$$
\Delta(w[x|y])
=
w \otimes w[x|y] +
w[x] \otimes wx[y] +
w[x|y] \otimes wxy,
$$
where $w,x,y \in \Pi$.
When we apply this to $\Pi = F$
we see that,
for arbitrary 1-cochains $u,v\colon \beta_1(F) \to g_{\phi}$,
their cup product $u \cup v$, evaluated
via the given $G$-invariant 2-form on $g$, amounts to the
2-cochain
which assigns
$$
(u \cup v)[x|y] = u(x) \cdot (\roman{Ad}(\phi(x))v(y))
$$
to $[x|y] \in B_2(F)$. \qed
\enddemo

The chosen 2-chain $c$ in $C_2(F)$ looks like
$$
c = \sum \nu_{j,k}[x_j|x_k].
$$
Define
the 2-form
$\omega_{c,\cup,\phi}$
on $Z^1(F,g_{\phi})$
by the explicit formula
$$
\omega_{c,\cup,\phi}(u,v) = \langle c, u \cup v \rangle
=\sum \nu_{j,k}
u(x_j) \cdot (\roman{Ad}(\phi(x_j))v(x_k)),
\quad
u,v \in Z^1(F,g_{\phi}).
$$
Notice this 2-form will {\it not\/}
be antisymmetric;
only its restriction to the space $Z^1(\pi,g_{\phi})$
of $g_{\phi}$-valued 1-cocycles for $\pi$
is antisymmetric.
By construction,
the
2-form $\omega_{\kappa,\cdot,\phi}$
is induced by the restriction of
$\omega_{c,\cup,\phi}$
to the space $Z^1(\pi,g_{\phi})$.
Hence Theorem 4.4 will be a consequence of the following.

\proclaim{Lemma 4.6}
At the point $\phi$ of $\roman{Hom}(F,G)$,
the 2-form $\omega_c$ is the antisymmetrization
of $\omega_{c,\cup,\phi}$.
\endproclaim

The 2-form $\omega_c$ on
$\roman{Hom}(F,G)$
arises from the 2-form $\Omega$ on $G^2$,
pulled back
via the evaluation map
$E$ from $F^2 \times \roman{Hom}(F,G)$  to $G^2$
and evaluated at the 2-chain $c$.
More precisely,
for $x,y \in F$ fixed,
consider
the smooth map
$\phi_{[x|y]}$
from $\roman{Hom}(F,G)$
to $G^2$
which assigns $(\phi(x),\phi(y))$ to
$\phi \in \roman{Hom}(F,G)$
and let
$
\omega_{[x|y]} = \phi_{[x|y]}^*(\Omega);
$
then
$$
\omega_c = \sum \nu_{j,k}
\omega_{[x_j|x_k]}.
$$
Thus it is manifest that Lemma 4.6 is a consequence of the following.

\proclaim{Lemma 4.7}
For every $x,y \in F$, the 2-form
$\omega_{[x|y]}$
is given by the formula
$$
\omega_{[x|y]}(u,v) = \frac 12 \left(
u(x)
\cdot (\roman {Ad}(\phi(x))v(y))
-
(\roman {Ad}(\phi(x))u(y)) \cdot
v(x)\right),
\tag4.7.1
$$
whatever
$u,v \in \roman T_{\phi}(\roman{Hom}(F,G))$,
the vector space $\roman T_{\phi}(\roman{Hom}(F,G))$ being identified
with the space of $g_{\phi}$-valued 1-cocycles
for $F$ via right translation.
In other words,
for every $x,y \in F$, the 2-form
$\omega_{[x|y]}$
amounts to
the antisymmetrization of the
2-form on
$\roman T_{\phi}(\roman{Hom}(F,G))$
given by the assignment
$$
(u,v) \longmapsto
\langle u\cup v,[x|y] \rangle.
$$
\endproclaim

\demo{Proof}
Writing $\roman T_{\phi} = \roman T_{\phi}\roman{Hom}(F,G)$,
we  must calculate the composite
$$
\roman T_{\phi}
\otimes
\roman T_{\phi}
@>>>
\left(\roman T_{\phi(x)}G \times \roman T_{\phi(y)}G\right)
\otimes
\left(\roman T_{\phi(x)}G \times \roman T_{\phi(y)}G\right)
@>{\omega_1 \otimes \overline \omega_2 -\overline \omega_2 \otimes\omega_1 }>>
g \otimes g
$$
and combine it with the given pairing.
In other words, we must compute
the two composites
$$
\align
\roman T_{\phi}
\otimes
\roman T_{\phi}
&@>>>
\roman T_{\phi(x)}G \otimes \roman T_{\phi(y)}G
@>{\omega \otimes \overline \omega}>>
g \otimes g
@>{\cdot}>> \bold R
\\
\roman T_{\phi}
\otimes
\roman T_{\phi}
&@>>>
\roman T_{\phi(y)}G
\otimes
\roman T_{\phi(x)}G
@>{\overline \omega\otimes\omega}>>
g \otimes g
@>{\cdot}>> \bold R .
\endalign
$$
However,
right translation
identifies
the space
$\roman T_{\phi}$
with the space of $g_{\phi}$-valued 1-cocycles for $F$.
Moreover a composite of the kind
$$
\roman T_{\phi}
@>>>
\roman T_{\phi(y)}G
@>{\overline \omega}>>
g
$$
amounts to the assignment
of the value $v(y)$
to a $g_{\phi}$-valued 1-cocycle
$v$ for $F$
since $\overline \omega$ is the canonical {\it right\/} invariant
$g$-valued 1-form on $G$;
likewise,
the adjoint representation $\roman {Ad}$ being viewed
as a $0$-form on $G$ with values in $\roman{Aut}(g)$,
we have $\overline \omega =\roman {Ad}\,\omega $ whence
a composite of the kind
$$
\roman T_{\phi}
@>>>
\roman T_{\phi(x)}G
@>{\omega}>>
g
$$
amounts to the assignment
of the value $\roman {Ad}(\phi(x^{-1}))u(x)$
to a $g_{\phi}$-valued 1-cocycle
$u$ for $F$.
Thus all told,
for
every $u,v \in \roman T_{\phi}$,
$$
\omega_{[x|y]}(u,v) = \frac 12
\left((\roman {Ad}(\phi(x^{-1}))u(x))
\cdot v(y)
-
u(y) \cdot
(\roman {Ad}(\phi(x^{-1}))v(x))\right).
$$
Since the 2-form is $G$-invariant,
this identity is equivalent to (4.7.1), and
the assertion follows. \qed
\enddemo

The proof of Theorem 4.4 is now complete.

\proclaim{Corollary 4.8}
Right translation identifies the
restriction of the
2-form $\omega_{c,\Cal P}$
to the kernel of
the derivative
$\roman T_{\phi} r$
with
the 2-form
on $\roman Z^1(\pi,g_{\phi})$
obtained as the
composite of
$\omega_{\kappa,\cdot,\phi}$
with the projection
from
$\roman Z^1(\pi,g_{\phi})$
to
$\roman H^1(\pi,g_{\phi})$.
Consequently
the alternating 2-form $\omega_{c,\Cal P}$
on
$\Cal H(\Cal P,G)$ induces
the 2-form
$\omega_{\kappa,\cdot,\phi}$
on
$\roman H^1(\pi,g_{\phi})$,
whatever $\Cal P$ and $c$.
\endproclaim

\beginsection 5. The case of a surface group

Let
$$
\Cal P = \langle x_1,y_1,\dots,x_\ell,y_\ell; r\rangle,\quad
r = \Pi [x_j,y_j],
$$
be the
standard presentation of the
fundamental group $\pi$ of a closed surface $\Sigma$
of genus $\ell$, and let
$\kappa \in \roman H_2(\pi)$
be a generator.
This implies
that the 2-chain $c \in C_2(F)$
has $\partial c = \pm r \in C_2(F)$,
and we assume the choices have been made in such a way that
$\partial c =  r$.
Further, suppose that the
given 2-form $\cdot$ on $g$ is non-degenerate.
Write $Z$ for the centre of $G$ and $z$ for its Lie algebra.
By Poincar\'e duality
in the cohomology of $\pi$, for every
$\phi$
in the pre-image $r^{-1}(z) \subseteq \Cal H(\Cal P,G)$
of $z \subseteq O$, in particular,
for every $\phi \in \roman{Hom}(\pi,G)$,
the 2-form
$\omega_{\kappa,\cdot,\phi}$
is then symplectic.
This form is just that considered by {\smc Goldman} \cite\goldmone.
Furthermore, the identity
$
\delta_G (\omega_{c,\Cal P})
= d \mu^{\sharp}
$
says that,
for every $X \in g$,
$$
\omega_{c,\Cal P}(X_{\Cal H(\Cal P,G)},\cdot\,)
= d (X \circ \mu),
$$
that is, formally the momentum mapping property is satisfied.
This together with
the symplecticity of the 2-form
$\omega_{\kappa,\cdot,\phi}$
at every
$\phi \in
r^{-1}(z)$
implies that
$\omega_{c,\Cal P}$
has maximal rank
equal to
$\dim \Cal H(\Cal P,G)$
at every point of
the pre-image
$r^{-1}(z)$ of $z$, in particular, at every point of
$\roman{Hom}(\pi,G)$.
In fact,
the symplecticity of the 2-form
$\omega_{\kappa,\cdot,\phi}$
at
$\phi \in r^{-1}(z)$
implies that
the 2-form $\omega_{c,\Cal P}$
on
the tangent space
$\roman T_{\phi}\Cal H(\Cal P,G)\cong\roman C^1(\Cal P, g_{\phi})$,
restricted to the subspace
$\roman Z^1(\pi,g_{\phi})$ of 1-cocycles,
has degeneracy space
equal to the subspace $\roman B^1(\pi,g_{\phi})$ of 1-coboundaries,
and the momentum mapping property then implies that
the 2-form $\omega_{c,\Cal P}$
on
the whole space
$\roman C^1(\Cal P, g_{\phi})$ is non-degenerate.
Let
$\Cal M(\Cal P,G)$
be the subspace of
$\Cal H(\Cal P,G)$
where
the 2-form $\omega_{c,\Cal P}$
is non-degenerate;
this is an open $G$-invariant
subset containing
the pre-image $r^{-1}(z)$.
In other words,
$\Cal M(\Cal P,G)$
is a smooth $G$-manifold and
the 2-form $\omega_{c,\Cal P}$ is in fact a $G$-invariant symplectic
structure on it.
Moreover, the restriction
$$
\mu = -\psi \circ r \colon
\Cal M(\Cal P,G)
@>>>
g^*
$$
is a momentum mapping
in the usual sense.
By the Corollary to Lemma 1,
the derivative of $\psi$ at the origin
equals  the adjoint of the given 2-form
which is assumed non-degenerate,
whence $\psi$ is regular near the origin
and the zero locus of
$\mu$ coincides with that of $r$, that is, with the space
$\roman{Hom}(\pi,G)$.
Symplectic reduction then yields the space
$\roman{Rep}(\pi,G)$.

\medskip\noindent{\bf 6. Twisted moduli spaces}\smallskip\noindent
The universal central extension
of $\pi$ arises from
the presentation $\Cal P$
in the following way:
Let $F$ be the free group on the generators,
$N$ the normal closure of $r$ in $F$,
and
$\Gamma = F\big / [F,N]$;
then the
kernel of the canonical projection from $\Gamma$ to
$\pi$ is a copy of the integers,
generated by
$[r] = r[F,N]\in \Gamma = F\big / [F,N]$, and
these combine to
the {\it universal central extension\/}
$$
0
@>>>
\bold Z
@>>>
\Gamma
@>>>
\pi
@>>>
1
$$
of $\pi$.
Write $Z$ for the centre of $G$,
let $z$ be the Lie algebra of $Z$, and
let $X \in z$.
When $G$ is connected, $z$ coincides with the zentre of $g$
but in general $z$ equals the invariants for the induced
action of the group $\pi_0$ of components of $G$ on the centre of $g$.
Let $\roman{Hom}_X(\Gamma,G)$
denote the space of homomorphisms
$\phi$ from $\Gamma$ to $G$
having the property that
$\phi[r] = \roman{exp}(X)$;
we assume $X$ chosen so that
$\roman{Hom}_X(\Gamma,G)$ is non-empty.
We comment on the significance of this assumption below.
Let $\roman{Rep}_X(\Gamma,G)=\roman{Hom}_X(\Gamma,G)\big / G$,
the resulting {\it twisted moduli space\/}
or {\it twisted representation space\/}.
The
choice of generators identifies
the space $\roman{Hom}_X(\Gamma,G)$ with the pre-image
of
$\roman{exp}(X) \in G$, for the word map
$r$ from
$G^{2\ell}$ to $G$
induced by the relator $r$, and we can play a similar
game as before, with the same choice
of $c \in C_2(F)$ so that $\partial c = r$
represents $\kappa \in \roman H_2(\pi)$.
More precisely,
since
the centre of $G$ is contained in $O$,
the space $\roman{Hom}_X(\Gamma,G)$
arises as
the pre-image of $X \in z \subseteq O$
under the map
$r$ from $\Cal H(F,G)$ to $O$.
Furthermore,
in view of the Corollary to Lemma 1,
the map $\psi$ from $g$ to $g^*$
is regular at every point of the centre of $g$,
in fact, the restriction of $\psi$ to the centre
equals  the adjoint of
the given 2-form
whence
the space $\roman{Hom}_X(\Gamma,G)$
equals
the pre-image of
the adjoint
$X^{\sharp}\in g^*$
of $X$
under the momentum mapping $\mu$
from $\Cal M(\Cal P,G)$ to $g^*$.
Consequently the space $\roman{Rep}_X(\Gamma,G)$
is the corresponding reduced space,
for the coadjoint orbit
in $g^*$
consisting of the single point $X^{\sharp}$.
\smallskip
We now explain briefly the geometric significance of the spaces
$\roman{Rep}_X(\Gamma,G)$ for $G$ compact:
Let $\Gamma_{\bold R} $ denote the non-connected Lie group
arising from $\Gamma$ when its centre
$\bold Z\langle [r]\rangle$
is extended to the reals $\bold R$.
A homomorphism $\phi$ from $\Gamma$ to $G$
having the property that
$\phi[r] = \roman{exp}(X)$ extends canonically to a homomorphism
$\Phi$ from $\Gamma_{\bold R} $ to $G$
by the assignment
$\Phi(x) = \phi(x)$ for $ x \in \Gamma$ and
$\Phi(t[r]) = \roman {exp} (tX) \in Z$,
for $t \in \bold R$.
This homomorphism, in turn,
determines a principal
$G$-bundle $\xi \colon P \to \Sigma$
together with a central Yang-Mills connection $A$ on $\xi$,
and this association in fact identifies
the moduli space $N(\xi)$ of central Yang-Mills connections
with a connected component of the space
$\roman{Rep}_X(\Gamma,G)$.
Details may be found
in \cite\atibottw\ for the case of connected $G$
and in our paper \cite\topology\ for the general case.
The element $X$ is in fact a topological characteristic class
of the corresponding bundle $\xi$.
However, when the structure group $G$
is not connected, it may happen that
the space $\roman{Rep}_X(\Gamma,G)$
is not connected
and
topologically inequivalent bundles
may still have the same characteristic class
$X$.
In particular we see that the requirement
that $X$ be chosen in such a way that
the space $\roman{Hom}_X(\Gamma,G)$
is non-empty is topological in nature.
Moreover, it is now clear that the space
of representations of $\Gamma$ in $G$ with the property that
the value of $[r]$ lies in the centre of $G$
is partitioned into a disjoint union
of representation spaces
of the kind
$\roman{Rep}_{X_\xi}(\Gamma,G)$,
where
$X_\xi \in z$ refers to the characteristic class
of a corresponding principal $G$-bundle $\xi$ on $\Sigma$.
The space
$\roman{Rep}(\pi,G)$
is the special case of this construction
for $X_\xi = 0$ and its  connected components
correspond to
topologically inequivalent flat $G$-bundles.
For example, for $G=\roman U(n)$,
the Lie algebra of the centre amounts to
$2\pi i \bold R$,
and the possible choices for $X$ are integral multiples
of $2\pi i$.
These are just the Chern classes of the corresponding bundles
$\xi$ --- notice a principal $\roman U(n)$-bundle
on a closed surface $\Sigma$
is topologically classified
by its Chern class in $\roman H^2(\pi,\pi_1(\roman U(n)))$.
For such a bundle $\xi$, the moduli space $N(\xi)$
is homeomorphic to
the {\smc Narasimhan-Seshadri}
\cite\narashed\
moduli space of semistable
rank $n$ holomorphic vector
bundles of degree equal to the Chern class of $\xi$.
Thus our construction in particular
yields these moduli spaces
by {\it symplectic reduction, applied to
a smooth finite dimensional symplectic
manifold with a hamiltonian action of
the finite dimensional Lie group\/} $\roman U(n)$.
In \cite\atibottw\ these spaces
have been obtained only by reduction applied to
the {\it infinite\/} dimensional space of all connections,
with hamiltonian action of the group of gauge transformations.

\beginsection 7. Applications

Suppose $G$ compact.
Recall that the notion of stratified symplectic space
has been introduced in \cite\sjamlerm.

\proclaim {Theorem 7.1}
With respect to the decomposition according to $G$-orbit types,
the space
$\roman{Rep}(\pi,G)$
and, more generally,
each
twisted representation space
$\roman{Rep}_X(\Gamma,G)$
inherits a structure of stratified symplectic space.
\endproclaim

In fact, the momentum mapping
$\mu$
from $\Cal M(\Cal P,G)$ to $g^*$
is proper;
this follows from the fact that
it arises from
the word mapping
$r$ from $G^{2\ell}$
to $G$, which is proper since
$G^{2\ell}$
and $G$ are compact.
The statement of the Theorem is thus an immediate consequence
of the main result of
\cite\sjamlerm.
The structure
of stratified symplectic space has been obtained in our paper
\cite\poisson\
by another method.

\proclaim{Theorem 7.2}
Each stratum of the space
$\roman{Rep}(\pi,G)$
and, more generally,
each stratum of a
twisted representation space
$\roman{Rep}_X(\Gamma,G)$
has finite symplectic volume.
\endproclaim

The proof follows the same pattern
as that
for the argument for
(3.9) in \cite\sjamlerm.
There the unreduced symplectic manifold is assumed compact.
However the compactness of the zero level set suffices;
in our situation the zero level set {\it is\/} compact.
In fact, it suffices to prove the statement for the local model
in \cite\sjamlerm\
which looks like the reduced space of a unitary representation
of a compact Lie group, for the corresponding unique momentum
mapping having the value zero at the origin.
For the local model there is no difference between
(3.9) in \cite\sjamlerm\
and our situation.
Once the statement is established for the local model,
that of Theorem 7.2
follows since the reduced space
may be covered by finitely many open sets
having a local model of the kind described.
\smallskip
We mention two other consequences:

\proclaim{Corollary 7.3}
There is a unique open, connected, and dense stratum.
\endproclaim

In fact, this follows at once from \cite\sjamlerm\ (5.9).
A different argument has been given in our paper \cite\singulat.
Likewise
\cite\sjamlerm\ (5.11) entails the following,
also observed in \cite\poisson.

\proclaim{Corollary 7.4}
The reduced Poisson algebra is symplectic, that is,
its only Casimir elements are the constants.
\endproclaim

\bigskip

\centerline{References}
\medskip

\ref \no  \atiboton
\by M. Atiyah and R. Bott
\paper The moment map  and equivariant cohomology
\jour Topology
\vol 23
\yr 1984
\pages  1--28
\endref

\ref \no  \atibottw
\by M. Atiyah and R. Bott
\paper The Yang-Mills equations over Riemann surfaces
\jour Phil. Trans. R. Soc. London  A
\vol 308
\yr 1982
\pages  523--615
\endref

\ref \no \bergever
\by N. Berline, E. Getzler, and M. Vergne
\book Heat kernels and Dirac Operators
\publ Springer
\publaddr Berlin $\cdot$ Heidelberg $\cdot$ New York $\cdot$ Tokyo
\yr 1992
\endref

\ref \no \bottone
\by R. Bott
\paper On the Chern-Weil homomorphism and the continuous cohomology of
Lie groups
\jour Advances
\vol 11
\yr 1973
\pages  289--303
\endref

\ref \no \botshust
\by R. Bott, H. Shulman, and J. Stasheff
\paper On the de Rham theory of certain classifying spaces
\jour Advances
\vol 20
\yr 1976
\pages 43--56
\endref

\ref \no \goldmone
\by W. M. Goldman
\paper The symplectic nature of the fundamental group of surfaces
\jour Advances
\vol 54
\yr 1984
\pages 200--225
\endref

\ref \no \singula
\by J. Huebschmann
\paper The singularities of Yang-Mills connections
for bundles on a surface. I. The local model
\paperinfo preprint February 1992
\endref

\ref \no \singulat
\by J. Huebschmann
\paper The singularities of Yang-Mills connections
for bundles on a surface. II. The stratification
\paperinfo preprint February 1992
\endref

\ref \no \topology
\by J. Huebschmann
\paper
Holonomies of Yang-Mills connections
for bundles on a surface with disconnected structure group
\paperinfo preprint 1992
\endref

\ref \no \smooth
\by J. Huebschmann
\paper
Smooth structures on
moduli spaces of central Yang-Mills connections
for bundles on a surface
\paperinfo preprint 1992
\endref

\ref \no \poisson
\by J. Huebschmann
\paper
Poisson
structures on certain
moduli spaces
for bundles on a surface
\paperinfo preprint 1993
\endref

\ref \no \locpois
\by J. Huebschmann
\paper Poisson geometry of flat connections
for {\rm SU(2)}-bundles on surfaces
\paperinfo preprint 1993
\endref

\ref \no \singulth
\by J. Huebschmann
\paper The singularities of Yang-Mills connections
for bundles on a surface. III. The identification of the strata
\paperinfo in preparation
\endref

\ref \no \jeffrone
\by L. Jeffrey
\paper
Extended moduli spaces of flat connections
on Riemann surfaces
\jour Math. Ann
\paperinfo to appear
\endref

\ref \no \jeffrtwo
\by L. Jeffrey
\paper Symplectic forms on moduli spaces
of flat connections on 2-manifolds
\paperinfo manuscript, preliminary version
\endref

\ref \no \karshone
\by Y. Karshon
\paper
An algebraic proof for the symplectic
structure of moduli space
\jour Proc. Amer. Math. Soc.
\vol 116
\yr 1992
\endref

\ref \no \kirwafou
\by F. Kirwan
\paper
Convexity properties of the moment mapping, {\rm III}
\jour Inventiones
\vol 77
\yr 1984
\pages 547--552
\endref

\ref \no \kirwaboo
\by F. Kirwan
\book Cohomology of quotients in symplectic and algebraic geometry
\publ Princeton University Press
\publaddr Princeton New Jersey
\yr 1984
\endref

\ref \no \maclaboo
\by S. Mac Lane
\book Homology
\bookinfo Die Grundlehren der mathematischen Wissenschaften
 No. 114
\publ Springer
\publaddr Berlin--G\"ottingen--Heidelberg
\yr 1963
\endref

\ref \no \narshtwo
\by M. S. Narashiman and C. S. Seshadri
\paper Holomorphic vector bundles on a compact Riemann surface
\jour Math. Ann.
\vol 155
\yr 1964
\pages  69--80
\endref

\ref \no \narashed
\by M. S. Narashiman and C. S. Seshadri
\paper Stable and unitary vector bundles on a compact Riemann surface
\jour Ann. of Math.
\vol 82
\yr 1965
\pages  540--567
\endref

\ref \no \sjamlerm
\by R. Sjamaar and E. Lerman
\paper Stratified symplectic spaces and reduction
\jour Ann. of Math.
\vol 134
\yr 1991
\pages 375--422
\endref

\ref \no \weinstwe
\by A. Weinstein
\paper On the symplectic structure of moduli space
\paperinfo preprint 1992; A. Floer memorial, Birkh\"auser, to appear
\endref

\enddocument